\documentclass[11pt]{article}
\usepackage{graphics,amsmath,amssymb,amsfonts}
\def \k{[\![}
\def \j{]\!]}
\def \ui#1{^{(#1)}}

\def \hh{\vskip0.5\baselineskip \hbox to \hsize}

\def \ds{\displaystyle\displaystyle}
\def \li{\ds\frac{1}{\lambda}}
\def \V#1{{\cal V}_{#1}}
\def \Q#1{{\cal Q}_{#1}(\lambda)}
\def \SQ#1#2{{\cal Q}^{#1}_{#2}(\lambda)}
\numberwithin{equation}{section}
\title{New symmetries for the Ablowitz-Ladik hierarchies }
\author{ Da-jun Zhang,$^{1}$\footnote{Corresponding author. E-mail: djzhang@staff.shu.edu.cn}
~~Tong-ke Ning,$^{2}$
~~Jin-bo Bi,$^{1}$
~~Deng-yuan Chen$^{1}$
\\
\small $^1${\it Department of Mathematics,
Shanghai University, Shanghai 200444,  P.R. China}\\
\small $^2${\it Science College, University of Shanghai for Science and Technology, Shanghai 200093, P.R. China}}

\begin{document}

\maketitle

\begin{abstract}

In the letter we give new  symmetries for the isospectral and non-isospectral
Ablowitz-Ladik hierarchies by means of the zero curvature representations
of evolution equations related to the Ablowitz-Ladik spectral problem.
Lie algebras constructed by symmetries are further obtained.
We also discuss the relations between the recursion operator
and isospectral and non-isospectral flows.
Our method can be generalized to other systems to construct
symmetries for non-isospectral equations.

\end{abstract}

\vskip 30pt

\section {Introduction}

It is well-known that infinitely many symmetries and their Lie algebra serve as one of
mathematical structures of integrability for evolution equations\cite{Fokas-symm-SAM-1987}.
In general, a Lax integrable isospectral evolution equation can have two sets of symmetries,
isospectral and non-isospectral symmetries, or called $K$- and $\tau$-symmetries, respectively.
One efficient way to construct $\tau$-symmetries was proposed by Fuchssteiner\cite{Fuchssteiner-MasSym-PTP-1983}
by using the master symmetry.
This method was later developed to many continuous
(1+1)-dimensional Lax integrable systems\cite{Chen-Zhang-JPA-1990, Chen-Zhang-JMP-1996},
(1+2)-dimensional systems\cite{Chen-Xin-Zhang-CSF-2003}
and further to some differential-difference cases\cite{Ma-Fuchssteiner-JMP-1999,
Ma-Tamizhimani-JPSJ-1999}.

This letter will discuss $K$- and $\tau$-symmetries  for the isospectral
Ablowitz-Ladik(AL) hierarchy, which is a well-known discrete
hierarchy\cite{Ablowitz-Ladik-JMP-1975}-\cite{Ablowitz-Ladik-SAM-1977}.
We will also construct new infinitely many symmetries for the
non-isospectral AL hierarchy.
The AL spectral problem can have two sets of isospectral
hierarchies\cite{Zeng-AL-JPA-1995} which respectively correspond to positive and negative powers of
the spectral parameter $\lambda$ in the time-evolution part in Lax pair. The same results
hold for the non-isospectral hierarchies as well, as shown in \cite{Ma-Tamizhimani-JPSJ-1999},
where the algebraic relations between isospectral and non-isospectral flows related to
positive powers of $\lambda$ and the algebraic relations between isospectral and non-isospectral flows
related to negative powers of $\lambda$ were discussed, respectively.

Our method to construct $K$- and $\tau$-symmetries for the isospectral AL hierarchy
is essentially the same as used in Ref.\cite{Ma-Tamizhimani-JPSJ-1999,Ma-Fuchssteiner-JMP-1999},
and as well as a direct generalization of its continuous version\cite{Chen-Zhang-JMP-1996}.
Recently, we uniformed the two sets of isospectral flows
(positive order and negative order)
to one hierarchy with a uniform recursion operator\cite{ZDJ-CDY-JPA-2002}.
This motivates us to do the same thing for the two sets of non-isospectral flows.
Then we investigate the algebraic relations of the uniformed isospectral flows and
uniformed non-isospectral flows.
As a result, we can generate new symmetries for those isospectral
AL evolution equations and get their Lie algebra.
And most important, we can construct infinitely many symmetries for the
non-isospectral AL hierarchy and derive their Lie algebra.
We also discuss the relations between the recursion operator
and isospectral and non-isospectral flows.

The letter is organized as follows. Sec.2 lists out some basic notations.
In Sec.3, we give the isospectral and non-isospectral
AL hierarchies and their zero curvature representations.
In Sec.4, we construct two sets of symmetries for the isospectral
AL hierarchy, give their Lie algebra
and discuss the relations between the recursion operator
and isospectral and non-isospectral flows.
In Sec.5, we construct  symmetries for the non-isospectral
AL hierarchy and give their Lie algebra.

\section{Basic notations}

To make our discussions smooth and convenient,
let us redescribe some notations in \cite{ZDJ-CDY-JPA-2002}.

Assume that
$U_{2}=\{ u_n \equiv u(t,n)=(u_{1}(t,n),u_{2}(t,n))^{T}\}$ is a
vector field space, where $\{ u_i(t,n)\}$  are all real   functions
defined over $\mathbb{R}\times \mathbb{Z}$ and  vanish rapidly as $|n|\rightarrow\infty$.
By $\V{2}$ denote a
linear space consisting of all vector fields
$f=(f_{1}(u(t,n)),f_{2}(u(t,n)))^{T}$ living on $U_2$.
where $\{ f_i(u(t,n))\} $ are $C^{\infty}$ differentiable with respect to $t$ and
$n$, $C^{\infty}$-Gateaux differentiable with respect to $u_n$, and
$f_i(u(t,n))|_{u_n=0}=0$. Then let $\Q{2}$ denote a Laurent matrix
polynomials space composed by all $2\times 2$ matrixes
$Q=Q(\lambda,u(t,n))=(q_{ij}(\lambda,u(t,n)))_{2\times 2}$, where
$\{ q_{ij}\}$ (or $Q$) are all the Laurent (matrix)  polynomials of
$\lambda$. Besides, we define  two subspaces of $\Q{2}$ as
\begin{equation}
\SQ{+}{2}= \{ Q\in \Q{2} |~{\rm the ~lowest~ degree ~of} ~
\lambda \geq 0\}
\end{equation}
\begin{equation}
\SQ{-}{2}= \{ Q\in \Q{2} |~{\rm the ~highest~ degree ~of}~
\lambda \leq 0\}.
\end{equation}

The Gateaux derivative of $f \in
\V{2}$ (or $f$ being an operator on $\V{2}$) in the direction $g \in
\V{2}$ is
defined  by
\begin{equation}
f' [g]=\left. \frac{d}{d\varepsilon}\right| _{\varepsilon=0}
f(u+\varepsilon g),
\end{equation}
and  the Lie product for any $f,g\in
\V{2}$ is described as
\begin{equation}
\k f,g\j =f' [g]-g'[f].
\end{equation}
Besides, for a given discrete evolution equation
${u_n}_t =K(u_n)$, $\sigma (u_n)\in \V{2}$ is called  its symmetry if
\begin{equation}
\sigma_t=K'[\sigma],
\end{equation}
or equivalently,
\begin{equation}
\frac{\partial\sigma}{\partial t}=\k K,\sigma \j.
\label{def-symmetry-2}
\end{equation}

\section {Isospectral and non-isospectral AL hierarchies}

The well-known AL spectral problem is given as\cite{Ablowitz-Ladik-JMP-1975}-\cite{Ablowitz-Ladik-SAM-1977}
\begin{equation}
E\phi=M\phi, ~~M=\Biggl (
                  \begin{array}{cc}
                   \lambda    &  Q_n   \\
                  R_n  & \frac{1}{\lambda}
                 \end{array}
             \Biggr ),~~
     u_n =\Biggl(
           \begin{array}{c}
               Q_n\\
               R_n
          \end{array}
        \Biggr ), ~~
    \phi= \Biggl (
           \begin{array}{c}
            \phi_1\\
            \phi_2
          \end{array}
         \Biggr ),
\label{AL-n}
\end{equation}
where $E$ is an shift operator defined as $E^j f(n)=f(n+j),~\forall j\in \mathbb{Z}$.
From the compatibility condition of \eqref{AL-n} and its corresponding time evolution
\begin{equation}
\begin{array}{l}
\phi_t=N\phi, ~N=
       \Biggl(
       \begin{array}{cc}
         A_n  & B_n\\
         C_n  & D_n
        \end{array}
      \Biggr), \end{array}
\label{AL-t}
\end{equation}
i.e.,
the zero-curvature equation
\begin{equation}
M_t=(EN)M-MN,
\label{zero-c-eq}
\end{equation}
one can easily get\cite{ZDJ-CDY-JPA-2002}
\begin{equation*}
    A_n=\frac{1}{\lambda} (E-1)^{-1}(-R_nEB_n +Q_nC_n)+\frac{n \lambda_t}{\lambda}+ a_0,
\end{equation*}
\begin{equation*}
    D_n=\lambda (E-1)^{-1}(R_nB_n -Q_nEC_n)-\frac{n \lambda_t}{\lambda} +d_0,
\end{equation*}
and
\begin{equation}
{u_n}_t=(\lambda L_1 -\li L_2)
            \left (\begin{array}{c}
                    B_n\\
                    C_n
                   \end{array}
            \right )
      +(a_0-d_0)\left (\begin{array}{c}
                    Q_n\\
                    -R_n
                   \end{array}
            \right )
      +\frac{(2n +1 )\lambda_t}{\lambda}\left (\begin{array}{c}
                    Q_n\\
                    -R_n
                   \end{array}
            \right ),
\label{u-t}
\end{equation}
where
$a_0=A_n|_{u_n=0}-\frac{n \lambda_t}{\lambda}$, $d_0=D_n|_{u_n=0}+\frac{n \lambda_t}{\lambda}$,
\begin{equation}
L_1= \left ( \begin{array}{cc}
                   -1  & 0\\
                   0   & E
                 \end{array}
         \right )
       +\left (\begin{array}{c}
                    -Q_n\\
                    R_nE
                   \end{array}
         \right )
         (E-1)^{-1}(R_n, -Q_n E),
\end{equation}
\begin{equation}
L_2= \left ( \begin{array}{cc}
                   -E  & 0\\
                   0   & 1
                 \end{array}
         \right )
       -\left (\begin{array}{c}
                    -Q_nE\\
                    R_n
                   \end{array}
         \right )
         (E-1)^{-1}(R_nE, -Q_n ).
\end{equation}

For the iosopectral case, i.e., $\lambda_t=0$, expanding $(B_n,C_n)^T$ in $\SQ{-}{2}$ and $\SQ{+}{2}$ respectively,
we can get two different sets of isospectral hierarchies\cite{Zeng-AL-JPA-1995,ZDJ-CDY-JPA-2002}.
Our method used here is little bit different from Ref.\cite{ZDJ-CDY-JPA-2002}.
Expanding
\begin{equation}
\Biggl(
    \begin{array}{c}
        B_{n}\\
        C_{n}
      \end{array}
    \Biggr)=\sum^{l}_{j=0}
\Biggl(
    \begin{array}{c}
        b^{-}_{n,j}\\
        c^{-}_{n,j}
      \end{array}
    \Biggr )\lambda^{-2(l-j)-1},~~~~(l \geq 0)
\label{expand-minus}
\end{equation}
and setting $(b^{-}_{n,0},c^{-}_{n,0})^T\equiv (0,0)^T$
and $a_0=-d_0=\frac{1}{2}\lambda^{-2l}$, from \eqref{u-t} we can get
\begin{equation*}
\begin{array}{l}
{u_n}_t=(1-\delta_{0,l})L_1
\Biggl(
    \begin{array}{c}
        b^{-}_{n,l}\\
        c^{-}_{n,l}
      \end{array}
    \Biggr )+\delta_{0,l}\Biggl (
    \begin{array}{c}
        Q_{n}\\
        -R_{n}
      \end{array}
    \Biggr ),
\\
L_1\Biggl(
    \begin{array}{c}
        b^{-}_{n,l-j}\\
        c^{-}_{n,l-j}
      \end{array}
    \Biggr)=L_2
\Biggl (
    \begin{array}{c}
        b^{-}_{n,l-j+1}\\
        c^{-}_{n,l-j+1}
      \end{array}
    \Biggr ),~~j=1,2,\cdots,l-1,
\\
L_2\Biggl(
    \begin{array}{c}
        b^{-}_{n,1}\\
        c^{-}_{n,1}
      \end{array}
    \Biggr)=
\Biggl (
    \begin{array}{c}
        Q_{n}\\
        -R_{n}
      \end{array}
    \Biggr ).
\end{array}
\end{equation*}
Then, taking $(b^{-}_{n,1},c^{-}_{n,1})^T=-(Q_{n-1},R_{n})^T$ yields a negative order isospectral hierarchy
\begin{equation}
{u_n}_t=K\ui{-l}=L^{-l} K\ui{0}, ~~l=0,1,2,\cdots,
\label{isosp-minus}
\end{equation}
where
\begin{equation}
{K}\ui{0}
 =\Biggl (
    \begin{array}{c}
        Q_{n}\\
        -R_{n}
      \end{array}
    \Biggr ),
\end{equation}
and the recursion operator $L$ is defined by
\begin{equation}
\begin{array}{ll}
     L= L_2L^{-1}_1
     = & \Biggl( \begin{array}{cc}
                   E  & 0\\
                   0  & E^{-1}
                 \end{array}
         \Biggr )
       +\Biggl (\begin{array}{c}
                    -Q_nE\\
                    R_n
                   \end{array}
         \Biggr)
         (E-1)^{-1}(R_nE, Q_n E^{-1})\\
          &~~~+\mu_n \Biggl(\begin{array}{c}
                    -EQ_{n}\\
                    R_{n-1}
                   \end{array}
         \Biggr )
         (E-1)^{-1}(R_n, Q_n) \ds\frac{1}{\mu_n}, ~~~~~(\mu_{n}=1-Q_nR_n).
   \end{array}
\label{L}
\end{equation}
Similarly, expanding $(B_n,C_n)^T$ in  $\SQ{+}{2}$ as
\begin{equation}
\Biggl (
    \begin{array}{c}
        B_{n}\\
        C_{n}
      \end{array}
    \Biggr )=\sum^{l}_{j=0}
\Biggl(
    \begin{array}{c}
        b^{+}_{n,j}\\
        c^{+}_{n,j}
      \end{array}
    \Biggr)\lambda^{2(l-j)+1},~~~~(l \geq 0),
\label{expand-plus}
\end{equation}
setting $(b^{+}_{n,0},c^{+}_{n,0})^T\equiv (0,0)^T$
and taking $a_0=-d_0=\frac{1}{2}\lambda^{2l}$ and $(b^{+}_{n,1},c^{+}_{n,1})^T=(Q_{n},R_{n-1})^T$,
we can get another isospectral hierarchy, i.e., a positive order hierarchy,
\begin{equation}
{u_n}_t=K\ui{l}=L^{l} K\ui{0}, ~~l=0,1,2,\cdots.
\label{isosp-plus}
\end{equation}
Thus, \eqref{isosp-minus} and \eqref{isosp-plus} can be uniformed to one hierarchy as\cite{ZDJ-CDY-JPA-2002}
\begin{equation}
{u_n}_t=K\ui{l}=L^{l} K\ui{0}, ~~l\in \mathbb{Z}.
\label{isosp-uniform}
\end{equation}
The recursion operator $L$ is a strong and hereditary symmetry operator for the above
hierarchy\cite{ZDJ-CDY-JPA-2002},
and this hierarchy has been shown to have multi-Hamiltonian structures\cite{Zeng-AL-JPA-1995,ZDJ-CDY-JPA-2002}
and infinitely many conservation laws\cite{ZDJ-CSF-2002}.

For the non-iosopectral case, i.e., $\lambda_t\neq 0$,
we first
expanding $(B_n,C_n)^T$ in $\SQ{-}{2}$ as \eqref{expand-minus},
setting $(b^{-}_{n,0},c^{-}_{n,0})^T\equiv (0,0)^T$
and taking $a_0=-d_0=0$, $\lambda_t=\lambda^{-2l+1}$
and $(b^{-}_{n,1},c^{-}_{n,1})^T=-((2n-1)Q_{n-1},(2n+1)R_{n})^T$, we can get the
negative order non-isospectral hierarchy
\begin{equation}
{u_n}_t=\sigma\ui{-l}=L^{-l} \sigma\ui{0}, ~~l=0,1,2,\cdots,
\label{non-isosp-minus}
\end{equation}
where
\begin{equation}
\sigma\ui{0}
 =(2n+1)\Biggl(
    \begin{array}{c}
        Q_{n}\\
        -R_{n}
      \end{array}
    \Biggr ).
\end{equation}
Expanding $(B_n,C_n)^T$ in  $\SQ{+}{2}$ as \eqref{expand-plus},
setting $(b^{+}_{n,0},c^{+}_{n,0})^T\equiv (0,0)^T$
and taking $a_0=-d_0=0$, $\lambda_t=\lambda^{2l+1}$,
and $(b^{+}_{n,1},c^{+}_{n,1})^T=((2n+1)Q_{n},(2n-1)R_{n-1})^T$,
we  get the positive order non-isospectral hierarchy
\begin{equation}
{u_n}_t=\sigma\ui{l}=L^{l} \sigma\ui{0}, ~~l=0,1,2,\cdots,
\label{non-isosp-plus}
\end{equation}
and these two non-isospectral hierarchies can be uniformed to one hierarchy as
\begin{equation}
{u_n}_t=\sigma\ui{l}=L^{l} \sigma\ui{0}, ~~l\in \mathbb{Z}.
\label{non-isosp-uniform}
\end{equation}

$\{K^{(l)}\}$ and $\{\sigma^{(l)}\}$ are called isospectral flows and non-isospectral flows, respectively.
On the basis of the above discussions, it is not difficult to give  the zero curvature representations of
these two sets of flows.

{\bf Proposition 3.1} ~{\it $\{K^{(l)}\}$ and $\{\sigma^{(l)}\}$ have the following
zero curvature representations
\begin{equation}
M'[K^{(l)}]=(EN^{(l)})M-MN^{(l)},
\label{z-c-p-isosp}
\end{equation}
\begin{equation}
M'[\sigma^{(l)}]=(EU^{(l)})M-MU^{(l)}-M_{\lambda}\lambda_t,
\label{z-c-p-non-isosp}
\end{equation}
where
\begin{equation}
\lambda_t=\lambda^{2l+1}, ~~~(l \in \mathbb{Z}),
\end{equation}
$K^{(l)},\sigma^{(l)}\in \mathcal{V}_2$,
$N^{(l)},U^{(l)} \in \Q{2}$ and satisfy
\begin{equation}
N^{(l)}|_{u_n=0}=\frac{1}{2}\lambda^{2l}
\Biggl(\begin{array}{cc}
1 & 0 \\
0 & -1
\end{array}
\Biggr ),~~~
U^{(l)}|_{u_n=0}=\lambda^{2l}
\Biggl(\begin{array}{cc}
n & 0 \\
0 & -n
\end{array}
\Biggr ),~~~ (l \in \mathbb{Z}).
\label{asymp-cond}
\end{equation}
}\hfill $\Box$

Besides, we have the following properties\cite{ZDJ-CDY-JPA-2002} on the AL spectral problem \eqref{AL-n}.

{\bf Proposition 3.2} ~{\it The following matrix equation
\begin{equation}
M'[X]=(EN)M-MN,~~~X\in \V{2},~N\in \mathcal{Q}_2(\lambda)~ {\rm and}~ N|_{u_n=0}=0
\label{cond-1}
\end{equation}
has only zero solutions $X=0$ and $N=0$.
}\hfill $\Box$

{\bf Proposition 3.3} ~{\it For any given $Y\neq 0 \in \V{2}$, there exist
solutions $N^\pm \in \mathcal{Q}_{2}^{\pm}(\lambda)$
satisfying
\begin{equation}
M' [L^{\pm 1}Y-\lambda^{\pm 2}Y]=(EN^\pm)M -MN^\pm ,~ ~ N^\pm|_{u_n=0}=0
\label{cond-2}
\end{equation}
where $L^{+1}$ denotes $L$ defined by \eqref{L}.
}\hfill $\Box$

\section{Symmetries for the isospectral AL hierarchy and Lie algebra}

In this section, we discuss the algebras of the flows $\{K^{(l)}\}$ and $\{\sigma^{(l)}\}$
and the related symmetries for the isospectral AL hierarchy.
We also give the relations between the recursion operator $L$
and these two sets of flows.

\subsection{Symmetries for the isospectral AL hierarchy and Lie algebra}

In this subsection, we will directly generate the method used in Ref.\cite{Chen-Zhang-JMP-1996}
to construct $K$- and $\tau$-symmetries for the isospectral AL hierarchy. Our method is also
essentially the same as used in Ref.\cite{Ma-Fuchssteiner-JMP-1999,Ma-Tamizhimani-JPSJ-1999}.

We list out our results through the following propositions.

{\bf Proposition 4.1} ~{\it If the isospectral and non-isospectral flows $\{K^{(l)}\}$ and $\{\sigma^{(l)}\}$
have their zero curvature representations \eqref{z-c-p-isosp} and \eqref{z-c-p-non-isosp},
then, $\forall l,s \in \mathbb{Z}$, the Lie products of these flows satisfy
\begin{equation}
\begin{array}{l}
M'[\k K^{(l)}, K^{(s)} \j ]=(E<N^{(l)},N^{(s)}>)M-M<N^{(l)},N^{(s)}>,
\\
M'[\k K^{(l)}, \sigma^{(s)} \j ]=(E<N^{(l)},U^{(s)}>)M-M<N^{(l)},U^{(s)}>,
\\
M'[\k \sigma^{(l)}, \sigma^{(s)} \j ]=(E<U^{(l)},U^{(s)}>)M-M<U^{(l)},U^{(s)}>
-2(s-l)M_{\lambda}\lambda^{2(l+s)+1},
\end{array}
\label{K-s}
\end{equation}
where
\begin{equation}
\begin{array}{l}
<N^{(l)},N^{(s)}>={N^{(l)}}'[ K^{(s)}]-{N^{(s)}}'[ K^{(l)}]+[N^{(l)},N^{(s)}],
\\
<N^{(l)},U^{(s)}>={N^{(l)}}'[ \sigma^{(s)}]-{U^{(s)}}'[ K^{(l)}]+[N^{(l)},U^{(s)}]
+N^{(l)}_{\lambda}\lambda^{2s+1},
\\
<U^{(l)},U^{(s)}>={U^{(l)}}'[ \sigma^{(s)}]-{U^{(s)}}'[ \sigma^{(l)}]+[U^{(l)},U^{(s)}]
+U^{(l)}_{\lambda}\lambda^{2s+1}-U^{(s)}_{\lambda}\lambda^{2l+1},
\end{array}
\label{N-U}
\end{equation}
and satisfy
\begin{equation}
\begin{array}{l}
<N^{(l)},N^{(s)}>|_{u_n=0}=0,
\\
<N^{(l)},U^{(s)}>|_{u_n=0}=2l N^{(l+s)}|_{u_n=0},
\\
<U^{(l)},U^{(s)}>|_{u_n=0}=2(l-s) U^{(l+s)}|_{u_n=0}.
\end{array}
\label{N-U-0}
\end{equation}
Here, $[A,B]=AB -BA$.
}

{\it Proof:}~ A similar proof procedure can be found in Ref.\cite{Chen-Zhang-JMP-1996}.
\eqref{K-s} and \eqref{N-U} can be derived from
the zero curvature representations \eqref{z-c-p-isosp} and \eqref{z-c-p-non-isosp}
by making use of the identity
\begin{equation}
M'[\k f, g \j ]=(M'[f])'[g]-(M'[g])'[f],~~~ \forall f,g \in \V{2}.
\label{identity-1}
\end{equation}
\eqref{N-U-0} can be obtained by using the asymptotic conditions \eqref{asymp-cond}.
\hfill $\Box$

Then, using Proposition 3.1 and 3.2, we can have the following result.

{\bf Proposition 4.2} ~{\it The isospectral and non-isospectral flows $\{K^{(l)}\}$ and $\{\sigma^{(l)}\}$
form a Lie algebra $\mathcal{F}$ through the Lie product $\k \cdot, \cdot \j$;
and $\forall l,s \in \mathbb{Z}$ they have the following relations
\begin{equation}
\begin{array}{l}
\k K^{(l)}, K^{(s)} \j =0,
\\
\k K^{(l)}, \sigma^{(s)} \j =2l K^{(l+s)}, \\
\k \sigma^{(l)}, \sigma^{(s)} \j =2(l-s)\sigma^{(l+s)}.
\end{array}
\label{K-s-Lie}
\end{equation}
}
\hfill $\Box$

Different from Ref.\cite{Ma-Tamizhimani-JPSJ-1999}, here we have expanded the
Lie product relations between $\{K^{(l)}\}$ and $\{\sigma^{(s)}\}$
for any integer subindices. This can result in some interesting relations. For example,
$\forall s \in \mathbb{Z}$, we have
\begin{equation*}
\begin{array}{l}
\k K^{(0)}, \sigma^{(l)} \j \equiv 0,~~~\k K^{(l)}, \sigma^{(0)} \j =2l K^{(l)}, \\
\k K^{(l)}, \sigma^{(-l)} \j =2l K^{(0)}, ~~~
\k \sigma^{(l)}, \sigma^{(0)} \j =2l \sigma^{(l)}, ~~~
\k \sigma^{(l)}, \sigma^{(-l)} \j =4l \sigma^{(0)}.
\end{array}
\end{equation*}
Besides, by virtue of the above proposition, for any equation in the isospectral hierarchy
\eqref{isosp-uniform}, it is not difficult to get two sets of symmetries and their Lie algebra.

{\bf Proposition 4.3} ~{\it The arbitrary member in the isospectral hierarchy
\eqref{isosp-uniform},
\begin{equation}
{u_{n}}_t=K^{(l)},~~~ \forall l\in \mathbb{Z},
\end{equation}
has the following two sets of symmetries,
\begin{equation}
\{K^{(s)}\}~~{\rm and}~~ \{\tau^{(l,s)}=2ltK^{(l+s)}+\sigma^{(s)}\},~~~~ s\in \mathbb{Z},
\label{symmetries}
\end{equation}
which we call $K$-symmetries and $\tau$-symmetries, respectively.
They form a Lie algebra $\mathcal{S}$ and have the relations
\begin{equation}
\begin{array}{l}
\k K^{(q)}, K^{(s)} \j =0,
\\
\k K^{(q)}, \tau^{(l,s)} \j =2q K^{(q+s)}, \\
\k \tau^{(l,q)}, \tau^{(l,s)} \j =2(l-s)\tau^{(l,q+s)}.
\end{array}
\label{symmetry-Lie}
\end{equation}
}
\hfill $\Box$

From the Lie product relations \eqref{K-s-Lie} and \eqref{symmetry-Lie},
it is not difficult to find the generators of the Lie algebras $\mathcal{F}$ and $\mathcal{S}$.

{\bf Proposition 4.4} ~{\it The Lie algebra $\mathcal{F}$ can be generated by
the following four generators,
\begin{equation*}
\sigma^{(2)}, ~~ \sigma^{(-2)}, ~~
\sigma^{(1)} ({\rm or}~\sigma^{(-1)}),~~K^{(1)} ({\rm or} ~K^{(-1)}).
\end{equation*}
The Lie algebra $\mathcal{S}$ can be generated by
the following four generators,
\begin{equation*}
\tau^{(l,2)}, ~~ \tau^{(l,-2)}, ~~
\tau^{(l,1)} ({\rm or}~\tau^{(l,-1)}),~~K^{(1)} ({\rm or} ~K^{(-1)}).
\end{equation*}
}\hfill $\Box$

\subsection{Relations between the recursion operator and flows}

In this subsection, we discuss the relations between the recursion operator and flows.

{\bf Proposition 4.5} ~{\it For any $l\in \mathbb{Z}$, the flows
$K^{(l)}$ and $\sigma^{(l)}$ and their recursion operator $L$ satisfy
\begin{equation}
L'[K^{(l)}]-[{K^{(l)}}',L]=0,
\label{L-K}
\end{equation}
\begin{equation}
L'[\sigma^{(l)}]-[{\sigma^{(l)}}',L]-2L^{l+1}=0.
\label{L-sigma}
\end{equation}
}

{\it Proof:}~ \eqref{L-K} has been  proved in Ref.\cite{ZDJ-CDY-JPA-2002}.
For \eqref{L-sigma}, we prove
\begin{equation}
(L'[\sigma^{(l)}]-[{\sigma^{(l)}}',L]-2L^{l+1})Y=0,~~~ \forall Y\in \V{2},~~\forall l\in \mathbb{Z},
\label{(L-sigma)Y}
\end{equation}
i.e.,
\begin{equation}
L\k \sigma^{(l)},Y\j - \k \sigma^{(l)},LY \j -2L^{l+1}Y=0,~~~ \forall Y\in \V{2},~~\forall l\in \mathbb{Z}.
\label{(L-sigma)Y-1}
\end{equation}

$\forall Y \in \V{2}$, in the light of Proposition 3.3,
there exists $N^{+}$ and $W^{(l)}$ in $\Q{2}$ such that
\begin{equation}
M' [LY-\lambda^2 Y]=(EN^{+})M -MN^{+},~ ~ N^{+}|_{u_n=0}=0,
\label{p-1}
\end{equation}
\begin{equation}
M' [L\k \sigma^{(l)},Y\j]=(EW^{(l)})M -MW^{(l)} +\lambda^{2}M'[\k \sigma^{(l)},Y\j],~ ~ W^{(l)}|_{u_n=0}=0.
\label{p-2}
\end{equation}
Meanwhile, using zero curvature representation of $\sigma^{(l)}$, i.e., \eqref{z-c-p-non-isosp},
and the identity \eqref{identity-1}, we  have
\begin{equation}
\begin{array}{rl}
M' [\k \sigma^{(l)},Y\j]=&(E{U^{(l)}}'[Y])M
+(E{U^{(l)}})M'[Y] -M'[Y]U^{(l)}-M{U^{(l)}}'[Y]\\
~&~~~ -\lambda^{2l+1}M_{\lambda}'[Y]
 -(M'[Y])'[\sigma^{(l)}].
\end{array}
\label{p-3}
\end{equation}
On the other hand, using the identity \eqref{identity-1}, \eqref{p-1} and
zero curvature representation \eqref{z-c-p-non-isosp},
we can have
\begin{equation}
\begin{array}{rl}
M' [\k \sigma^{(l)},LY\j]\! =\!&\Bigl[E({U^{(l)}}'[LY]-{N^{+}}'[\sigma^{(l)}]+[U^{(l)},N^{+}]
-\lambda^{2l+1}N^{+}_{\lambda})\Bigr ]M\\
~&
-M\Bigl({U^{(l)}}'[LY]-{N^{+}}'[\sigma^{(l)}]+[U^{(l)},N^{+}]-\lambda^{2l+1}N^{+}_{\lambda}\Bigr )\\
~&
+\lambda^2\Bigl [(E{U^{(l)}})M'[Y] -M'[Y]U^{(l)}-\!\lambda^{2l}M'[Y]
- \lambda^{2l+1}M_{\lambda}'[Y]
-(M'[Y])'[\sigma^{(l)}]\Bigr ].
\end{array}
\label{p-4}
\end{equation}
Then \eqref{p-4} and \eqref{p-2} together with \eqref{p-3} yield
\begin{equation}
M' [L\k \sigma^{(l)},Y\j-\k \sigma^{(l)},LY\j-2\lambda^{2(l+1)} Y]
=(E<W^{(l)},U^{(l)},N^{+}>)M -M<W^{(l)},U^{(l)},N^{+}>,
\label{p-5}
\end{equation}
where
\begin{equation}
<W^{(l)},U^{(l)},N^{+}>=W^{(l)}-{U^{(l)}}'[LY-\lambda^{2}Y]+ {N^{+}}'[\sigma^{(l)}]
+[N^{+},U^{(l)}]+\lambda^{2l+1}N^{+}_{\lambda}
\end{equation}
satisfying
\begin{equation}
<W^{(l)},U^{(l)},N^{+}>|_{u_n=0}=0.
\end{equation}
Then, noting that \eqref{p-1} implies that there exists $\widetilde{N}^{+}\in \Q{2}$ such that
\begin{equation}
M' [L^{l+1}Y-\lambda^{2(l+1)} Y]=(E\widetilde{N}^{+})M -M\widetilde{N}^{+},~ ~ \widetilde{N}^{+}|_{u_n=0}=0,
\end{equation}
and using Proposition 3.2,
we can finally reach the equality \eqref{(L-sigma)Y-1} and thus we complete the proof.
\hfill $\Box$

We note that the algebra relations \eqref{K-s-Lie} can also be obtained through
the reductive approach by using \eqref{L-K} and \eqref{L-sigma}.

\section{Symmetries for the non-isospectral AL hierarchy and Lie algebra}

By virtue of the Lie product relations given in Proposition 4.2, we can construct
infinitely many symmetries for any member in the non-isospectral AL hierarchy
\eqref{non-isosp-uniform}.

{\bf Proposition 5.1}~ {\it For any $l\in \mathbb{Z}$, the non-isospectral evolution equation
\begin{equation}
{u_n}_t=\sigma\ui{l}
\label{non-isosp-l}
\end{equation}
has the following infinitely many symmetries
\begin{equation}
\eta^{(l,m)}=\sum^{m}_{j=0}C^{j}_{m}(2lt)^{m-j}\sigma^{(l-jl)}, ~~~ (m=0,1,2,\cdots),
\label{eta-symmetry}
\end{equation}
\begin{equation}
\gamma^{(l,m)}=\sum^{m}_{j=0}C^{j}_{m}(2lt)^{m-j}K^{(-jl)}, ~~~ (m=0,1,2,\cdots),
\label{gamma-symmetry}
\end{equation}
where $C^{j}_{m}=\frac{m!}{j!(m-j)!}$. For convenient, we call
\eqref{eta-symmetry} and \eqref{gamma-symmetry} the $\eta$-symmetries and
$\gamma$-symmetries, respectively.
\hfill $\Box$
}

This proposition can be proved by direct verification according to the definition
\eqref{def-symmetry-2} and the algebraic relations \eqref{K-s-Lie}.

{\bf Proposition 5.2}~ {\it $\eta$-symmetries $\{\eta^{(l,m)} \}_{m=0,1,2}$ and
$\gamma$-symmetries $\{\gamma^{(l,m)} \}_{m=0,1,2,\cdots}$ construct a Lie algebra $\widetilde{\mathcal{S}}$
and they follow the following Lie product relations,
\begin{equation}
\begin{array}{ll}
\k \eta^{(l,m)}, \eta^{(l,m)} \j =0, & ~(m=0,1,2),\\
\k \eta^{(l,m)}, \eta^{(l,s)} \j =2(s-m)l \eta^{(l,s+m-1)}, & ~(m,s=0,1,2,\cdots, ~m\neq s),\\
\k \gamma^{(l,m)}, \gamma^{(l,s)} \j =0, & ~(m,s=0,1,2,\cdots),\\
\k \eta^{(l,m)}, \gamma^{(l,0)} \j =0, & ~(m=0,1,2,\cdots),\\
\k \eta^{(l,m)}, \gamma^{(l,s)} \j =2 sl \gamma^{(l,s+m-1)}, & ~(m=0,1,2,\cdots, ~s=1,2,\cdots).
\end{array}
\label{eat-gamma}
\end{equation}
Obviously, the Lie  algebra $\widetilde{\mathcal{S}}$ has three generators
\begin{equation}
\eta^{(l,0)},~\eta^{(l,3)},~\gamma^{(l,1)}.
\end{equation}
}

{\it Proof:}~ We only prove the second and the last equalities in \eqref{eat-gamma}.
From \eqref{eta-symmetry} we have
\begin{equation*}
\begin{array}{rl}
\k \eta^{(l,m)}, \eta^{(l,s)} \j &
= \ds\ds \sum^{m}_{j=0}\sum^{s}_{h=0}
C^{j}_{m}C^{h}_{s}(2lt)^{m+s-j-h} \k \sigma^{(l-jl)}, \sigma^{(l-hl)} \j\\
~& =2l \ds\ds \sum^{m}_{j=0}\sum^{s}_{h=0}
C^{j}_{m}C^{h}_{s}(2lt)^{m+s-j-h} (h-j) \sigma^{(l-(j+h-1)l)}\\
~& =2l \ds\ds \sum^{m+s}_{i=1}\sum^{min\{i,m\}}_{j=0}
C^{j}_{m}C^{i-h}_{s}(2lt)^{m+s-i} (i-2j) \sigma^{(l-(i-1)l)}.
\end{array}
\end{equation*}
Without loss of generality, we let $m>s$.
Then, by noting that
\begin{equation}
\sum^{min\{i,m\}}_{j=0}
C^{j}_{m}C^{i-h}_{s}(2lt)^{m+s-i} (i-2j)=(s-m)C^{i-1}_{m+s-1},
~~(m,s=0,1,\cdots,~~m> s,~~1\leq i \leq m+s),
\label{comb-1}
\end{equation}
we immediately get
\begin{equation*}
\k \eta^{(l,m)}, \eta^{(l,s)} \j
=2(s-m)l \sum^{m+s-1}_{i=0}C^{i}_{m+s-1} \sigma^{(l-il)}
=2(s-m)l \eta^{(l,s+m-1)}.
\end{equation*}
Similarly, we can prove the last equality in \eqref{eat-gamma},
where we need to use the identity
\begin{equation}
\sum^{min\{i,m\}}_{j=0}
C^{j}_{m}C^{i-h}_{s}(2lt)^{m+s-i} (i-j)=s C^{i-1}_{m+s-1},
~~(m,s=0,1,\cdots,~~m> s,~~1\leq i \leq m+s).
\label{comb-2}
\end{equation}
The proof for \eqref{comb-1} and \eqref{comb-2} will be given in Appendix.
\hfill $\Box$

In addition, for the isospectral equation ${u_n}_t=K\ui{-l}$ and
non-isospectral equation ${u_n}_t=\sigma\ui{l}$,
they have a non-trivial mutual symmetry,
\begin{equation}
\sigma=-2ltK^{(0)}-K^{(-l)}+ \sigma^{(l)}.
\end{equation}


\vskip 12pt

\noindent{\bf Conclusion} ~

To sum up, in this letter, we first respectively
uniformed the isospectral AL hierarchy and non-isospectral AL hierarchy.
Then we derived Lie algebraic relations of these uniformed
flows by means of their zero curvature representations,
and consequently we obtained $K$-symmetries and $\tau$-symmetries
for the isospectral AL hierarchy.
As the Lie product relations between $\{K^{(l)}\}$ and $\{\sigma^{(s)}\}$
have been expended for any integer subindices $l$ and $s$,
some obtained symmetries are new.
And, as an important result, we worked out new infinitely many symmetries for the
non-isospectral AL hierarchy and gave their Lie algebra.
Generators of these obtained Lie algebras have been given.
The relations between the recursion operator $L$
and the two sets of flows were also discussed.
It is known that it is not easy to construct infinitely many symmetries for
non-isospectral evolution equations.
We believe that our method to derive symmetries for non-isospectral equations
through constructing negative order hierarchies is general and
can apply to other systems. This will be investigated in detail elsewhere.
In fact, there have been some known systems with inverse recursion operators,
i.e., with positive and negative order hierarchies\cite{QZJ-Strampp-2002}-\cite{SYP-CDY-XXX-2006}.
In addition, can our new symmetries lead to new reductions and solutions?

\vskip 16 pt
\noindent{\bf Acknowledgments}~

This project is supported by the National Natural Science
Foundation of China (10371070), the Youth Foundation of Shanghai Education Committee
and the  Foundation of Shanghai
Education Committee for Shanghai Prospective
Excellent Young Teachers.


\vskip 20pt

\leftline{\bf Appendix: Proof for \eqref{comb-1} and \eqref{comb-2} }

First, expanding both sides of the identity
$$(1+x)^m(1+x)^s=(1+x)^{m+s}$$
and picking up the coefficients of $x^i$ on both sides,
we can easily get
$$\sum^{min\{ m, i \}}_{j=0}
C^{j}_{m}C^{i-j}_s=C^{i}_{m+s},
~~~(m,s=0,1,2,\cdots,~~ 0\leq i\leq m+s).
\eqno(A.1)$$
Similarly, expanding both sides of the identity
$$\frac{d(1+x)^m}{d x}(1+x)^s=m(1+x)^{m+s-1}$$
and picking up the coefficients of $x^{i-1}$ on both sides
yield
$$\sum^{min\{ m, i \}}_{j=0}
j C^{j}_{m}C^{i-j}_s  = m C^{i-1}_{m+s-1},
~~~(s=0,1,\cdots,~~m=1,2,\cdots,~~ 1\leq i\leq m+s).
\eqno(A.2)$$

Next, by noting that $i C^{i}_{m+s}=(m+s)C^{i-1}_{m+s-1}$,
from $(A.1)$ we have
$$\sum^{min\{ m, i \}}_{j=0}
 i C^{j}_{m}C^{i-j}_s  = (m+s) C^{i-1}_{m+s-1},
~~~(s=0,1,\cdots,~~m=1,2,\cdots,~~ 1\leq i\leq m+s).
\eqno(A.3)$$
It then follows from $(A.2)$ and $(A.3)$ that
$$\sum^{min\{ m, i \}}_{j=0}
 (i-2j) C^{j}_{m}C^{i-j}_s  = (s-m) C^{i-1}_{m+s-1},
~~~(s=0,1,\cdots,~~m=1,2,\cdots,~~ 1\leq i\leq m+s)
\eqno(A.4)$$
and
$$\sum^{min\{ m, i \}}_{j=0}
 (i-j) C^{j}_{m}C^{i-j}_s  = s C^{i-1}_{m+s-1},
~~~(s=0,1,\cdots,~~m=1,2,\cdots,~~ 1\leq i\leq m+s),
\eqno(A.5)$$
i.e., \eqref{comb-1} and \eqref{comb-2}.

\end{document}